\documentclass[showpacs,oneside,twocolumn,pra,amsmath,amssymb,superscriptaddress]{revtex4-1}

\usepackage{cases}
\usepackage{amsmath}
\usepackage{amssymb}
\usepackage{amsfonts}
\usepackage{amssymb}
\usepackage{dcolumn}
\usepackage{bm}
\usepackage{bbm}
\usepackage{graphicx}
\usepackage{xcolor}
\usepackage{array}
\usepackage{subfigure}
\usepackage{hyperref}

\hypersetup{colorlinks=true,
            breaklinks=true,
            pdfstartview=Fit,
            linkcolor=blue,
            citecolor=blue,
            urlcolor=blue}

\begin{document}

\title{Experimental Test of Born's Rule by Inspecting Third-Order Quantum Interference on a Single Spin in Solids}

\author{Fangzhou Jin}
\author{Ying Liu}
\author{Jianpei Geng}
\affiliation{Hefei National Laboratory for Physical Sciences at the Microscale and Department of Modern Physics, University of Science and Technology of China, Hefei, 230026, China}
\author{Pu Huang}
\affiliation{Hefei National Laboratory for Physical Sciences at the Microscale and Department of Modern Physics, University of Science and Technology of China, Hefei, 230026, China}
\affiliation{Synergetic Innovation Center of Quantum Information and Quantum Physics, University of Science and Technology of China, Hefei, 230026, China}
\author{Wenchao Ma}
\affiliation{Hefei National Laboratory for Physical Sciences at the Microscale and Department of Modern Physics, University of Science and Technology of China, Hefei, 230026, China}
\author{Mingjun Shi}
\author{Chang-Kui Duan}\altaffiliation{ckduan@ustc.edu.cn}
\author{Fazhan Shi}
\author{Xing Rong}\altaffiliation{xrong@ustc.edu.cn}
\author{Jiangfeng Du}\altaffiliation{djf@ustc.edu.cn}
\affiliation{Hefei National Laboratory for Physical Sciences at the Microscale and Department of Modern Physics, University of Science and Technology of China, Hefei, 230026, China}
\affiliation{Synergetic Innovation Center of Quantum Information and Quantum Physics, University of Science and Technology of China, Hefei, 230026, China}

\begin{abstract}

 As a fundamental postulate of quantum mechanics, Born's rule assigns probabilities to the measurement outcomes of quantum systems and excludes multi-order quantum interference. Here we report an experiment on a single spin in diamond to test Born's rule by inspecting the third-order quantum interference. The ratio of the third-order quantum interference to the second-order in our experiment is ceiled at the scale of $10^{-3}$, which provides a stringent constraint on the potential breakdown of Born's rule.

\end{abstract}

\pacs{03.65.Ta, 03.65.-w, 76.30.Mi}
\maketitle

\section{Introduction}
In quantum mechanics, a measurement acting on a quantum system yields probabilistic outcomes that obey Born's rule (BR) \cite{Born.ZP1926}. BR is a fundamental postulate connecting the mathematical formalism of quantum theory with experiment \cite{book2009}.
It is typically formulated as \cite{Schlosshauer.FP2005}: if an observable $\hat{O}$, with eigenstates \{$|o_i \rangle$\} and spectrum \{$o_i$\}, is measured on a quantum system described by the state vector $|\psi\rangle$, the probability for the measurement to yield the value $o_i$ is given by $|\langle o_i|\psi\rangle|^2$.
There are theoretical attempts to encompass quantum mechanics as a special case \cite{Dakic.NJP2014}. Some may lead to the predictions deviating from that of BR  \cite{Sorkin.PLA1994}, for example, multi-order quantum interference vanishes according to BR but could exist in some generalized probabilistic theories. It is important to experimentally inspect multi-order quantum interference to test BR, since any significant nonzero observation of such interference would imply that BR did not strictly hold.
If BR were violated, the paradigm of quantum theory might need to be amended, and computational complexity of some problems could be reduced \cite{Aaronson.PRSLA2005}.

Sinha \emph{et al.} reported the first experimental measurement of the third-order interference term in an optical system to test BR \cite{Sinha.S2010}, and the ratio of the magnitude of the three-path interference to the expected two-path interference was bounded to less than $10^{-2}$.
Three-slit interference was used in their experiment. However, it has recently been shown \cite{Raedt.PRA2012} that the three-slit experiment involved the assumption of superposition principle. The assumption does not hold strictly, although it can be considered as a good approximation. This viewpoint was endorsed by Sinha \emph{et al.} in their latest theoretical works \cite{Sawant.PRL2014,Sinha.SP2015}. On the other hand, the detector size and position set significant limitations to the normalization scheme in three-slit experiment \cite{Gagnon.PRA2014}.
Thereafter, some other experiments utilizing interference of photons \cite{Hickmann.EL2011,Sollner.FP2012,Kauten.arXiv2015} and liquid-state nuclear magnetic resonance of a spin ensemble \cite{Park.NJP2012} were reported. While this NMR experiment is based on pseudo-pure states and the description of NMR system can be via the classical Bloch equations which do not involve BR. To verify BR experimentally,  one has to carry out experimental measures on pure states of a quantum system. A decisive experimental verification of BR with individual spin system is still lacking. Since individual spin systems have been widely used for quantum computation \cite{Ladd.N2010} and quantum metrology \cite{Balasubramanian.N2008,Maze.N2008,Grinolds.NP2013,Muller.NC2014,Shi.S2015} that involve BR, it is of practical significance to verify this rule on such systems.

In this paper, we perform an experiment to test BR by inspecting the third-order quantum interference in a nitrogen-vacancy (NV) center in diamond \cite{Gruber.S1997,Jelezko.PRL2004,Jelezko.PSSA2006,Dohertya.PR2013}.
NV center systems are convenient to initialize and readout \cite{Gruber.S1997}, has long coherence time \cite{Balasubramanian.NM2009,Maurer.S2012}, and can be manipulated with high precision \cite{Rong.PRL2014,Rong.nc2015}.
These advantages enable NV centers to be widely applied in quantum metrology \cite{Balasubramanian.N2008,Maze.N2008,Grinolds.NP2013,Muller.NC2014,Shi.S2015}, quantum computation \cite{Wrachtrup.J2006,Sar.N2012,Shi.PRL2010,Arroyo.NC2014}, and fundamental physics \cite{Waldherr.PRL2011,Hensen.N2015}.
In contrast to previous experiments using spatial paths of photons, our experiment harnessing energetic states of the single electron in an NV center provides an essential complement for testing BR.

\section{Testing Born's rule by a qutrit}
\begin{figure}
\centering
\includegraphics[width=0.85\columnwidth]{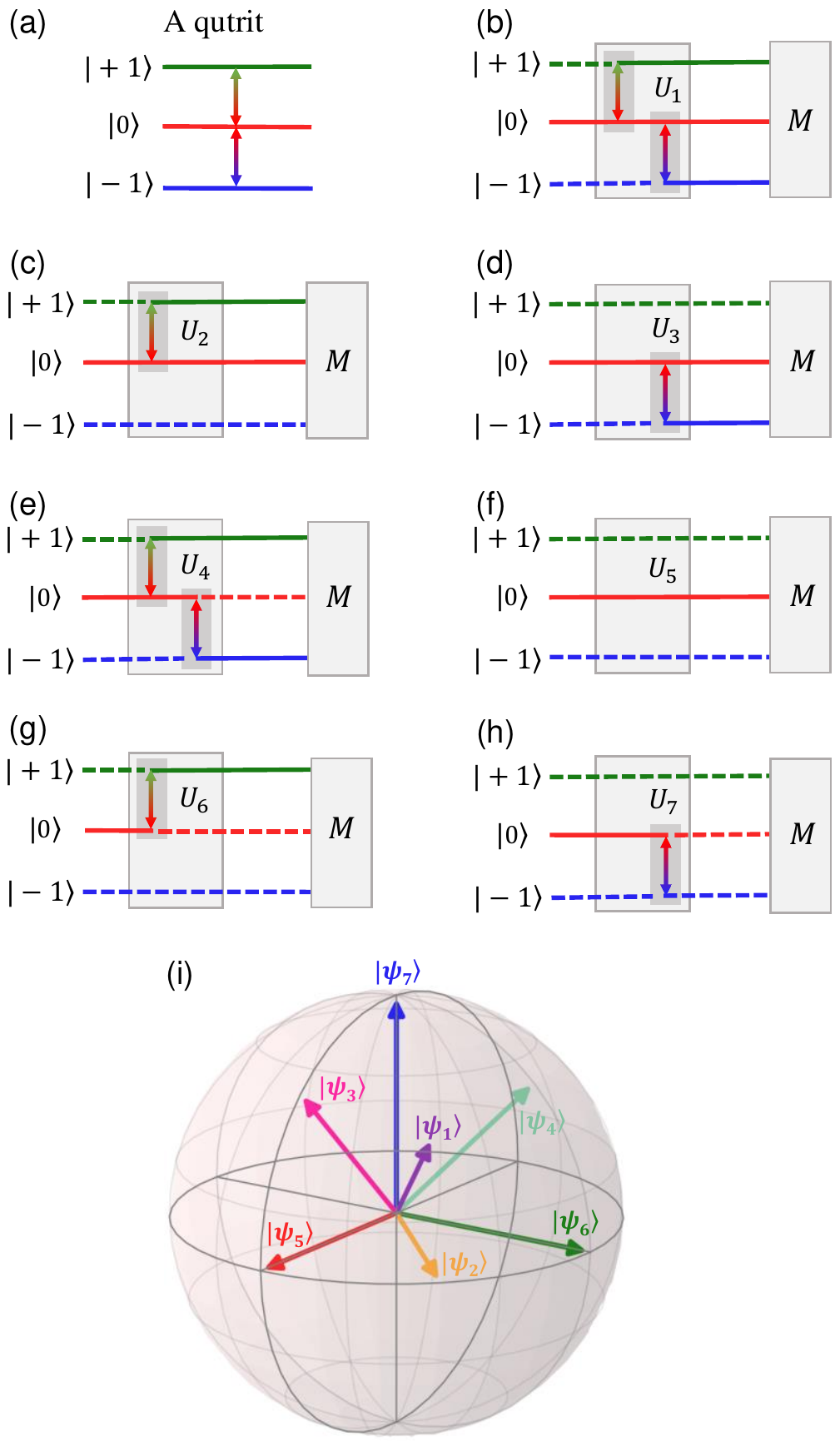}
\caption{(color online). Experimental processes for measuring the third-order quantum interference.
(a) A qutrit with a set of orthonormal basis $|0\rangle$, $|+1\rangle$ and $|-1\rangle$, with $|0\rangle\leftrightarrow|+1\rangle$ and $|0\rangle\leftrightarrow|-1\rangle$ being allowed transitions. (b)-(h) Preparation of the seven states \{$|\psi_i\rangle$\} ($i=1,2,\cdots,7$) by different operator $U_i$.
Measurement operator $M$ is acted on $|\psi_i\rangle$ in each of the seven experiments. The third-order quantum interference term can be obtained from the seven experiments.
(i) Representation of the state vectors \{$|\psi_i\rangle$\} in the case where $a=1/\sqrt{3}$ and $b=c=-1/\sqrt{3}$ in real three-dimensional space. The states \{$|\psi_i\rangle$\}, in turn, correspond to the prepared states in (b)-(h).
}
\label{fig1}
\end{figure}

Now we consider a qutrit with a set of orthogonal basis $|0\rangle$, $|+1\rangle$, and $|-1\rangle$, as shown in Fig.~\ref{fig1}(a). The qutrit is initialized to the state $|\psi_0\rangle=|0\rangle$, and then an operation $U_1$ prepares the state $|\psi_1\rangle=a |0\rangle+b |+1\rangle+c |-1\rangle$, as shown in Fig.~\ref{fig1}(b). After that, measurement operator $M=|m\rangle \langle m|$ is acted on the system, with an outcome $m$. Here $|m\rangle=\alpha |0\rangle+\beta |+1\rangle+\gamma |-1\rangle$, and $a$, $b$, $c$, $\alpha$, $\beta$, $\gamma$ are the probability amplitudes of quantum states. If BR holds, the probability of obtaining the measurement outcome $m$ is
\begin{equation} \label{p1}
\begin{split}
p_1&=\langle \psi_1|M|\psi_1\rangle=|\langle m|\psi_1\rangle|^2\\
% &=|\alpha|^2|a|^2+|\beta|^2|b|^2+|\gamma|^2|c|^2\\
% &~~~+(\alpha {\beta}^{\ast}{a}^{\ast}b+{\alpha}^{\ast} \beta a {b}^{\ast} )+(\alpha {\gamma}^{\ast}{a}^{\ast}c\\
% &~~~+{\alpha}^{\ast} \gamma a {c}^{\ast} )+(\beta {\gamma}^{\ast}{b}^{\ast}c+{\beta}^{\ast} \gamma b {c}^{\ast} )\\
 &=q_a+q_b+q_c+I_{ab}+I_{ac}+I_{bc},
 \end{split}
\end{equation}
where $q_a=|\alpha|^2|a|^2$, $q_b=|\beta|^2|b|^2$, $q_c=|\gamma|^2|c|^2$ and
$I_{ab}=\alpha {\beta}^{\ast}{a}^{\ast}b+{\alpha}^{\ast} \beta a {b}^{\ast}$, $I_{ac}=\alpha {\gamma}^{\ast}{a}^{\ast}c+{\alpha}^{\ast} \gamma a {c}^{\ast}$, $I_{bc}=\beta {\gamma}^{\ast}{b}^{\ast}c+{\beta}^{\ast} \gamma b {c}^{\ast}$. The subscripts $a$, $b$, and $c$ represent the terms related to $|0\rangle$, $|+1\rangle$, and $|-1\rangle$ respectively. The terms $I_{ab}$, $I_{ac}$, and $I_{bc}$ are regarded as the second-order quantum interference terms.  Similar to Ref.~\cite{Sinha.S2010}, we define the third-order interference term as the deviation of $p_1$ from the sum of the individual probabilities and the second-order interference terms:
\begin{equation} \label{Iabc}
I_{abc}^{(3)}:=p_1-(q_a+q_b+q_c+I_{ab}+I_{ac}+I_{bc}).
\end{equation}
According to BR, there is no third-order interference, i.e., $I_{abc}^{(3)}=0$.
In Eq.~(\ref{Iabc}), $p_1$ can be obtained in this experiment, while $q_a$, $q_b$, $q_c$ and $I_{ab}$, $I_{ac}$, $I_{bc}$ can be extracted from other experiments. In the following, we elaborate on how to measure these terms.

The initial state $|\psi_0\rangle$ followed by an operation $U_2$ yields the state $|\psi_2\rangle=(a |0\rangle+b |+1\rangle)/\sqrt{|a|^2+|b|^2}$, and then the measurement operator $M$ is acted on $|\psi_2\rangle$, as illustrated in Fig.~\ref{fig1}(c). The probability of obtaining the measurement outcome $m$ is
\begin{equation} \label{p2}
p_2=\langle \psi_2|M|\psi_2\rangle =\frac{1}{|a|^2+|b|^2}(q_a+q_b+I_{ab}).
\end{equation}

The same goes for the states $|\psi_3\rangle=(a |0\rangle+c |-1\rangle)/\sqrt{|a|^2+|c|^2}$, $|\psi_4\rangle=(b |+1\rangle+c |-1\rangle)/\sqrt{|b|^2+|c|^2}$, $|\psi_5\rangle=a /|a| |0\rangle$, $|\psi_6\rangle=b/|b||+1\rangle$, and $|\psi_7\rangle=c/|c||-1\rangle$, as shown in Fig.~\ref{fig1}(d)-(h). The Representation of the state vectors \{$|\psi_i\rangle$\} in the case where $a=1/\sqrt{3}$ and $b=c=-1/\sqrt{3}$ in real three-dimensional space are illustrated by Fig.~\ref{fig1}(i).
The probabilities of obtaining the measurement outcome $m$ are $p_3$, $p_4$, $p_5$, $p_6$, and $p_7$, respectively.
Therefore, the third-order interference term $I_{abc}^{(3)}$ can be written as
\begin{equation} \label{Iabc2}
\begin{split}
&I_{abc}^{(3)}=p_1-(|a|^2+|b|^2)p_2-(|a|^2+|c|^2)p_3\\
&-(|b|^2+|c|^2)p_4+|a|^2 p_5+|b|^2 p_6+|c|^2 p_7.
\end{split}
\end{equation}

The probabilities in Eq.~(\ref{Iabc2}), namely $p_1$, $p_2$, $p_3$, $p_4$, $p_5$, $p_6$, and $p_7$, can be obtained by the experiments.
The coefficients $a$, $b$, and $c$ are the probability amplitudes of the first state $|\psi_1\rangle$ which have been determined by the operation $U_1$. Since the goal is to test BR, the rule should not be presumed in preparing the states \{$|\psi_i\rangle$\}. Here \{$|\psi_i\rangle$\} can be prepared by different operation $U_i$ ($i=1,2,\cdots,7$), which do not involve BR. The detailed processes are shown in the Appendices.

To quantify the relative departure from BR, we define a normalized variant as
\begin{equation} \label{kappa}
\kappa=I_{abc}^{(3)}/I_{abc}^{(2)},
\end{equation}
where $I_{abc}^{(2)}=|I_{ab}|+|I_{ac}|+|I_{bc}|$
is the sum of the absolute values of the second-order interference terms, and $\kappa$ can be regarded as the ratio of an unexpected third interference term to the expected second interference term. A nonzero $I_{abc}^{(2)}$ ensures that it is in a quantum mechanical regime. The zero third-order interference term $I_{abc}^{(3)}$ means zero $\kappa$, and nonzero observation of the ratio $\kappa$ would violate BR.

\section{Experiment}
\begin{figure}
\centering
\includegraphics[width=0.85\columnwidth]{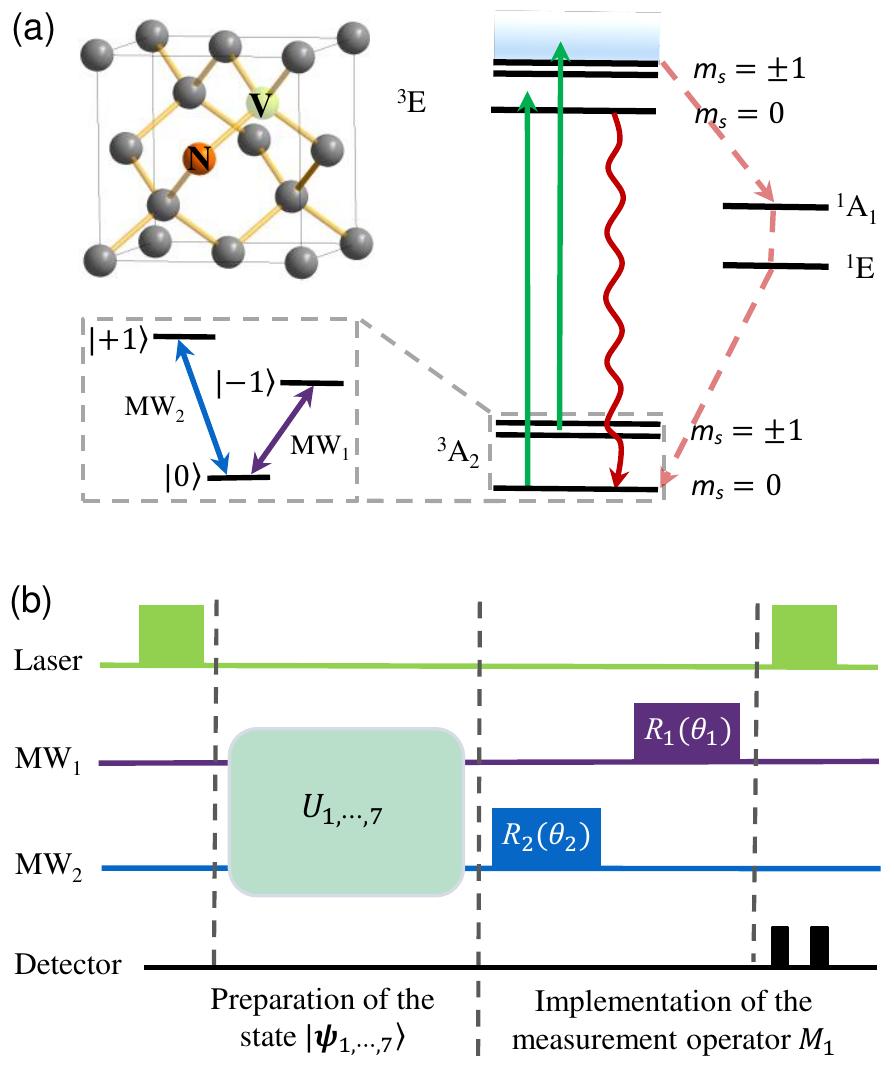}
\caption{(color online). Experimental system and pulse sequences. (a) An NV center in diamond is formed by a substitutional nitrogen atom and an adjacent vacancy. Spin projections $m_s=0,\pm1$ are defined with respect to the NV symmetry axis. Electronic spin polarization and readout is performed by optical excitation and red fluorescence detection. The microwave MW$_1$ and MW$_2$ drive the selective transitions  $|0\rangle\leftrightarrow|-1\rangle$ and $|0\rangle\leftrightarrow|+1\rangle$, respectively. (b) Each experiment includes two steps: preparation of the state $|\psi_i\rangle$ by $U_i$, and implementation of the measurement operator $M_1$. The rotation operations  $R_1(\theta_1)$ and $R_2(\theta_2)$ are realized by MW$_1$ and MW$_2$, respectively.
}
\label{fig2}
\end{figure}

We carry out the experiment on the electron spin of an NV center. As a defect in diamond, the NV center is composed of one substitutional nitrogen atom and an adjacent vacancy, as shown in Fig.~\ref{fig2}(a). The electronic ground state of the negatively charged NV center forms a spin triplet which is polarized by optical excitation, manipulated coherently by oscillating magnetic field, and readout by illuminating it again and collecting the state-dependent fluorescence \cite{Jelezko.PRL2004,Jelezko.PSSA2006,Dohertya.PR2013}.

The electronic ground states of an NV center in an external magnetic field parallel to the symmetry axis is described by the Hamiltonian $H=D S_z^2+\gamma_{\text{e}} B S_z,$
where the first term is zero-field splitting with $D=2.87$ GHz, the second term is Zeeman splitting with $\gamma_{\text{e}}=2.80$ MHz/G, $S_z$ is the spin angular momentum operator, and $B$ is the magnitude of the magnetic field.
In our experiment, the static magnetic field is along the NV symmetry axis with the magnitude $B \approx 510$ G. In this magnetic field, both the electron spin and the host nitrogen nuclear spin can be polarized by optical pumping. The energy levels with $m_s= 0, +1$, and $-1$ are labeled by $|0\rangle$, $|+1\rangle$, and $|-1\rangle$, respectively. Quantum states $|0\rangle$, $|+1\rangle$, and $|-1\rangle$ are chosen as the set of orthogonal basis. As shown in Fig.~\ref{fig2}(a), two channels of microwave MW$_1$ and MW$_2$ are resonant with the transitions $|0\rangle\leftrightarrow|-1\rangle$ and $|0\rangle\leftrightarrow|+1\rangle$, respectively. The rotation around $y$ axis (defined as the direction of the microwave field) by an angle $\phi$ with MW$_1$ and MW$_2$ are expressed as $R_1(\phi)$ and $R_2(\phi)$, respectively. The dephasing times $T_2^{\ast}$ is about $1.5~\mu$s. The count of photons emitted by the NV center and then collected by the avalanche photodiode was above 400K per second with the background below 5K per second.

\begin{figure}
\centering
\includegraphics[width=1\columnwidth]{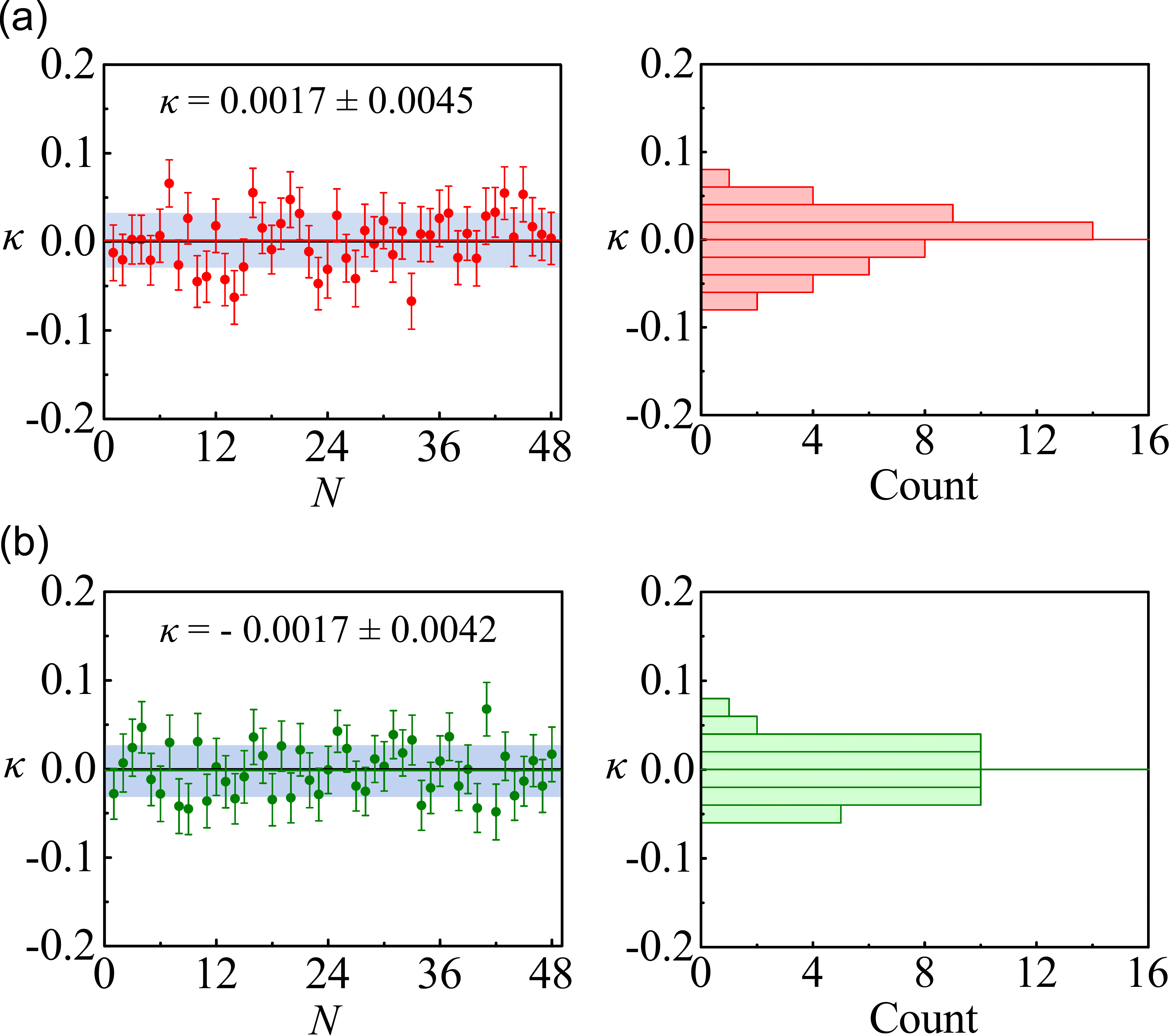}
\caption{(color online). Experimental data of the normalized variant $\kappa$. (a) The data associated with the measurement operator $M_1$ results in $\kappa=0.0017\pm0.0045$. (b) The data associated with the measurement operator $M_2$ results in $\kappa=-0.0017\pm0.0042$. The left figures are the values of $\kappa$ with errorbars representing standard deviations, and each data point is averaged by $2\times10^6$ times. The right figures are the distribution of $\kappa$ corresponding to the left. In left figures, the horizontal color lines represent the mean values of $\kappa$, and the blue shaded regions represent a band of one standard deviation of the distribution of $\kappa$ values around the mean.
}
\label{fig3}
\end{figure}

Fig.~\ref{fig2}(b) shows the pulse sequence for the experiments. The NV center is first initialized to the state $|0\rangle$ by a green laser pulse, and then prepared to the target state $|\psi_i\rangle$ by the corresponding operation $U_i$. The operation $U_i$ implemented by two microwave pulses MW$_1$ and MW$_2$ is given by $U_i=R_2(\phi_2^i) R_1(\phi_1^i)$. The rotation angles in the seven experiments are set to: $\phi_1^1=\arccos (1/3),~ \phi_2^1=\pi/2$; $\phi_1^2= \pi/2, ~\phi_2^2=0$; $\phi_1^3=0, ~ \phi_2^3=\pi/2$; $\phi_1^4=\pi/2, ~\phi_2^4=\pi$; $\phi_1^5=0, ~\phi_2^5=0$; $\phi_1^6=0,~ \phi_2^6=\pi$; $\phi_1^7=\pi, ~\phi_2^7=0$. These operations result in target states \{$|\psi_i\rangle\}$ with $a=1/\sqrt{3}$ and $b=c=-1/\sqrt{3}$. After $|\psi_i\rangle$ is prepared, the measurement operator $M_1$ is acted on the system. Here $M_1$ is implemented by the operations $R_2(\theta_2)$ and $R_1(\theta_1)$, and the final state selective detection \cite{Arroyo.NC2014}. The rotation angles $\theta_1$ and $\theta_2$ can be set to obtain the target measurement operator $M_1=|m_1\rangle \langle m_1|$ with $ |m_1\rangle =R_2^{\dag}(\theta_2) R_1^{\dag}(\theta_1) |0\rangle$. Then probability of the outcome is expressed as $p_i=|\langle m_1 |\psi_i\rangle|^2$.
We set $\theta_1=\theta_2=\pi /2$, so that $|m_1\rangle=1/2|0\rangle+1/2|+1\rangle+1/\sqrt{2}|-1\rangle$.
In the experiment, we measure the ratio of the photon counting for signal to that for reference \cite{Rong.PRL2014}. Here the measured ratio for the measurement operator $M_1$ acting on $|\psi_i\rangle$ is represented by $p_i'$. This ratio $p_i'$ is proportional to the probability $p_i$, i.e., $p_i' \propto p_i$, with $i=1,2,\cdots,7$. Hence the measured third-order interference term $I_{abc}^{(3)}{'}$, obtained from Eq.~(\ref{Iabc2}) with $p_i'$ instead of $p_i$, is also proportional to $I_{abc}^{(3)}$.
We measure each $p_i'$ for the seven states, and then the normalized ratio $\kappa$ can be obtained.

The experimental results are shown in Fig.~\ref{fig3}. We evaluate the normalized variant $\kappa$ from the experiments with the measurement operator $M_1$, and the result is $\kappa=0.0017\pm0.0045$, as shown in Fig.~\ref{fig3}(a). According to BR, $\kappa$ should be independent of measurement operators. We also evaluate $\kappa$ with another measurement operator $M_2$ by setting different rotation angles with $\theta_1=3\pi /2$ and $\theta_2=\pi /2$. The result is $\kappa=-0.0017\pm0.0042$ as depicted in Fig.~\ref{fig3}(b). Both results are in accordance with BR within the experimental errors. Higher precision for the measurement of $\kappa$ can be achieved by using an NV center with longer dephasing time \cite{Balasubramanian.NM2009,Maurer.S2012} and composite pulses with higher-fidelity \cite{Rong.PRL2014,Rong.nc2015}.

\section{Conclusion}
To summarize, we have experimentally tested BR by ruling out the third-order quantum interference using a single spin in diamond. We evaluated the ratio of third-order quantum interference to the second-order by using two different measurement operators, and bounded the ratios both at the scale of $10^{-3}$.  Our result provides a stringent constraint on the breakdown of BR.
This method can also be generalized to inspect higher-order quantum interference by employing the electron spin coupled with the host $^{14}$N nuclear spin.
The unification of quantum mechanics and gravitation may require the generalization of quantum mechanics, and this generalization probably involves the modification of BR. Our experiments as well as other experiments of this kind may be beneficial to setting limits on the extent of modification.

\section*{ACKNOWLEDGMENTS}
This work was supported by the National Key Basic Research Program of China (Grant No.\ 2013CB921800 and No. 2016YFB0501603), the National Natural Science Foundation of China (Grant No.\ 11227901, No.\ 31470835, No.\ 11274299 and No.\ 11275183) and the Strategic Priority Research Program (B) of the CAS (Grant No. XDB01030400). F.S. and X.R. thank the Youth Innovation Promotion Association of Chinese Academy of Sciences for the support.

\section*{APPENDIX A: Evolution of quantum states} \label{Appendix A}

The evolution of the quantum state $| \psi \rangle$ of a quantum system is described by the Schr\"{o}dinger equation
\begin{equation} \label{SchrodingEquation}
i\hbar \frac{\partial}{\partial t} | \psi \rangle = H | \psi \rangle,
\end{equation}
where $H$ is the Hamiltonian of the system and $\hbar$ is the reduced Planck constant.
Such evolution can be encapsulated into the time-evolution operator $U(t,t_0)$
\begin{equation} \label{timeevolution}
| \psi (t) \rangle = U(t,t_0) | \psi (t_0) \rangle.
\end{equation}
These two equations yield the Schr\"{o}dinger equation for the time-evolution operator
\begin{equation} \label{Schrodingoperator}
i\hbar \frac{\partial}{\partial t} U(t,t_0) = H U(t,t_0).
\end{equation}
From Eq.~(\ref{Schrodingoperator}) and the initial condition $U(t_0,t_0)=\mathbbm1$, one can obtain $U^{\dag}(t,t_0) U(t,t_0)=\mathbbm1$. It means that $U(t,t_0)$ is a unitary operator. In other words, the Schr\"{o}dinger equation ensures the unitary evolution of the quantum state. This  unitarity keeps the expression $ \langle \psi (t)| \psi (t) \rangle $ constant.
Up to now, Born's rule (BR) has not come into play.
In 1926, Max Born put forward BR in the context of scattering theory. Then the square of the amplitude modulus is explained as the probability of the measurement outcome.

In the main text, the quantum states \{$|\psi_i\rangle$\} are obtained from the same state $|\psi_0\rangle$ by different operation $U_i$. In order to avoid the assumption of BR, the initial state can be written in form of $|\psi_0\rangle=A |0\rangle$, where $A$ is a constant. According to the Schr\"{o}dinger equation, each of final quantum states satisfies the condition that the sum of the square of probability amplitudes modulus keeps constant $|A|^2$. This constant $|A|^2$ is a common coefficient in Eqs.~(1-4) of the main text, and it will be eliminated in the normalized variant $\kappa$. For simplicity, we set $A=1$ in the whole text.

\section*{APPENDIX B: Preparation of quantum states \{$|\psi_i\rangle$\}} \label{Appendix B}

As the goal is to test BR by ruling out third-order quantum interference, BR should not be presumed in preparing the states \{$|\psi_i\rangle$\}.
Now we give a detailed explanation of the preparation of \{$|\psi_i\rangle$\}.

The orthogonal basis $|0\rangle$, $|+1\rangle$, and $|-1\rangle$ can be represented by
\begin{equation} \label{kets}
|0\rangle=\begin{pmatrix}
0 \\
1 \\
0
\end{pmatrix},~
|+1\rangle=\begin{pmatrix}
1 \\
0 \\
0
\end{pmatrix},~
|-1\rangle=\begin{pmatrix}
0 \\
0 \\
1
\end{pmatrix}.
\end{equation}
Consider a spin-1 system such as the electronic ground states of an NV center, where $|0\rangle\leftrightarrow|+1\rangle$ and $|0\rangle\leftrightarrow|-1\rangle$ are allowed transitions, as shown in Fig.~1(a) of main text. With the application of an external microwave (MW) field around the $y$ axis, the Hamiltonian is written as
\begin{equation} \label{Hamiltonian}
H=H_0+H_t,
\end{equation}
with
\begin{equation} \label{Hamiltonian1}
H_0=D S_z^2+\gamma_{\text{e}} B S_z,~
H_t=\sqrt{2} \omega_1 \cos(\omega t)S_y,
\end{equation}
where $\omega$ and $\omega_1$, in turn, are the frequency of the MW and the Rabi frequency of the electron spin driven by the MW.
In the interaction picture, the interacting term of Hamiltonian is written as $H_I=e^{i H_0 t}H_t e^{-i H_0 t}$.
The external microwave field MW$_1$ and MW$_2$ drive the selective transitions $|0\rangle \leftrightarrow |-1\rangle$ and $|0\rangle \leftrightarrow |+1\rangle$, respectively. Under the rotating wave approximation, the corresponding time-evolution operators are given by
\begin{equation} \label{U1}
R_1(t)=\begin{pmatrix}
 1 & 0 & 0 \\
 0 & \cos \frac{\omega_1 t}{2} & \sin \frac{\omega_1 t}{2} \\
 0 & - \sin \frac{\omega_1 t}{2} & \cos \frac{\omega_1 t}{2} \\
\end{pmatrix},
\end{equation}
\begin{equation} \label{U2}
R_2(t)=\begin{pmatrix}
 \cos \frac{\omega_1 t}{2} & - \sin \frac{\omega_1 t}{2} & 0 \\
 \sin \frac{\omega_1 t}{2} & \cos \frac{\omega_1 t}{2} & 0 \\
 0 & 0 & 1 \\
\end{pmatrix}.
\end{equation}

The system is initialized to the state $|\psi_0\rangle=|0\rangle$.
For the first step, $|\psi_1\rangle$ is prepared by applying two operations, i.e.,
\begin{equation} \label{phi1}
|\psi_1\rangle  =R_2(t_{1}') R_1(t_1) |0\rangle\\=
\begin{pmatrix}
 -\cos \frac{\omega_1 t_1}{2} \sin \frac{\omega_1 t_{1}'}{2} \\
 \cos \frac{\omega_1 t_1}{2} \cos \frac{\omega_1 t_{1}'}{2} \\
 -\sin \frac{\omega_1 t_1}{2} \\
\end{pmatrix}.
\end{equation}
That means the coefficients $a, b$, and $c$ in main text are given by: $a=\cos \frac{\omega_1 t_1}{2} \cos \frac{\omega_1 t_{1}'}{2}$,
$b=-\cos \frac{\omega_1 t_1}{2} \sin \frac{\omega_1 t_{1}'}{2}$, and $c= -\sin \frac{\omega_1 t_1}{2}$, respectively. In this process, $t_1$, $t_{1}'$, and $\omega_1$ can be selected arbitrarily. Without loss of generality, $\omega_1$ is kept the same in the whole processes.

Then $|\psi_2\rangle$ is prepared by applying the operation $R_2(t_{2}')$, i.e.,
\begin{equation} \label{phi2}
|\psi_2\rangle  =R_2(t_{2}') |0\rangle=
\begin{pmatrix}
 -\sin \frac{\omega_1 t_{2}'}{2} \\
 \cos \frac{\omega_1 t_{2}'}{2} \\
 0 \\
\end{pmatrix}.
\end{equation}
In this process, if $t_{2}'$ is kept to satisfy the condition
\begin{equation} \label{condition2}
\frac{-\sin \frac{\omega_1 t_{2}'}{2}}{\cos \frac{\omega_1 t_{2}'}{2}}=\frac{-\cos \frac{\omega_1 t_1}{2} \sin \frac{\omega_1 t_{1}'}{2}}{\cos \frac{\omega_1 t_1}{2} \cos \frac{\omega_1 t_{1}'}{2}},
\end{equation}
$|\psi_2\rangle$ will be in form of $|\psi_2\rangle=(a |0\rangle+b |+1\rangle)/\sqrt{|a|^2+|b|^2}$. Then from Eq.~(\ref{condition2}), the parameter $t_{2}'$ is determined by $t_1$ and $t_{1}'$.

In the same way,  $|\psi_3\rangle$ is prepared by applying the operation $R_1(t_{3})$, i.e.,
\begin{equation} \label{phi3}
|\psi_3\rangle  =R_1(t_{3}) |0\rangle=
\begin{pmatrix}
 0 \\
\cos \frac{\omega_1 t_3}{2}  \\
 -\sin \frac{\omega_1 t_3}{2} \\
\end{pmatrix}.
\end{equation}
In this process, if $t_{3}$ is kept to satisfy the condition
\begin{equation} \label{condition3}
\frac{\cos \frac{\omega_1 t_3}{2}}{-\sin \frac{\omega_1 t_3}{2}}=\frac{\cos \frac{\omega_1 t_1}{2} \cos \frac{\omega_1 t_{1}'}{2} }{ -\sin \frac{\omega_1 t_1}{2}},
\end{equation}
$|\psi_3\rangle$ will be in form of $|\psi_3\rangle=(a |0\rangle+c |-1\rangle)/\sqrt{|a|^2+|c|^2}$. Similarly, the parameter $t_{3}$ is determined by $t_1$ and $t_{1}'$.

In the same way, $|\psi_4\rangle$ is prepared by applying the operations $R_1(t_{4})$ and $R_2(t_{4}')$, i.e.,
\begin{equation} \label{phi4}
|\psi_4\rangle  =R_2(t_{4}') R_1(t_4) |0\rangle\\=
\begin{pmatrix}
 -\cos \frac{\omega_1 t_4}{2} \sin \frac{\omega_1 t_{4}'}{2} \\
 \cos \frac{\omega_1 t_4}{2} \cos \frac{\omega_1 t_{4}'}{2} \\
 -\sin \frac{\omega_1 t_4}{2} \\
\end{pmatrix}.
\end{equation}
In this process, if $t_{4}$ and $t_{4}'$ are kept to satisfy the conditions
\begin{equation} \label{condition3}
\cos \frac{\omega_1 t_{4}'}{2}=0, ~
\frac{\cos \frac{\omega_1 t_4}{2}}{\sin \frac{\omega_1 t_4}{2}}=\frac{ \cos \frac{\omega_1 t_1}{2} \sin \frac{\omega_1 t_{4}'}{2} }{ \sin \frac{\omega_1 t_1}{2}},
\end{equation}
then $|\psi_4\rangle$ will be in form of $|\psi_4\rangle=(b |+1\rangle+c |-1\rangle)/\sqrt{|b|^2+|c|^2}$. Similarly, the parameter $t_{4}$ and  $t_{4}'$ are determined by $t_1$ and $t_{1}'$.

Besides, $|\psi_5\rangle$ is obtained without any operation, $|\psi_6\rangle$ is prepared by applying the operation $R_2(t_{6}')$ with $\sin \frac{\omega_1 t_{6}'}{2} =1$, and $|\psi_7\rangle$ is prepared by applying the operation $R_1(t_{7})$ with $\sin \frac{\omega_1 t_7}{2} =1$.

Consequently, the quantum states \{$|\psi_i\rangle$\} are obtained by different operations. In these processes, when the parameters $t_1$, $t_{1}'$, and $\omega_1$ are selected, the other parameters are determined as well.
In the experiment, it just needs to measure the duration of $\pi$ pulse with  Rabi frequency $\omega_1$. Then the other operations are obtained by applying pulses with corresponding time determined by $t_1$, $t_{1}'$, and $\omega_1$. The  duration of $\pi$ pulse can be measured by the Rabi oscillation, and this procedure does not involve the detailed form of BR.
In our experiments, the parameters $t_1$, $t_{1}'$ and $\omega_1$ are set as $ \frac{\omega_1 t_1}{2}=\arccos(1/3)$ and $\frac{\omega_1 t_{1}'}{2} =\pi/2$. Then the other parameters are determined as $ \frac{\omega_1 t_{2}'}{2}=\pi/2$, $ \frac{\omega_1 t_3}{2}=\pi/2$, $ \frac{\omega_1 t_{4}}{2}=\pi/2$, $\frac{\omega_1 t_{4}'}{2} =\pi$, $\frac{\omega_1 t_{6}'}{2} =\pi$, and $ \frac{\omega_1 t_{7}}{2}=\pi$.

\end{document}